\documentstyle[12pt]{article}

\begin{document}

\newcommand{\Om}{\Omega}
\newcommand{\df}{\stackrel{\rm def}{=}}
\newcommand{\co}{{\scriptstyle \circ}}
\newcommand{\de}{\delta}
\newcommand{\lb}{\lbrack}
\newcommand{\rb}{\rbrack}
\newcommand{\rn}[1]{\romannumeral #1}
\newcommand{\msc}[1]{\mbox{\scriptsize #1}}
\newcommand{\dsp}{\displaystyle}
\newcommand{\scs}[1]{{\scriptstyle #1}}

\newcommand{\ket}[1]{| #1 \rangle}
\newcommand{\bra}[1]{| #1 \langle}
\newcommand{\vac}{| \mbox{vac} \rangle }

\newcommand{\e}{\mbox{{\bf e}}}
\newcommand{\va}{\mbox{{\bf a}}}
\newcommand{\bc}{\mbox{{\bf C}}}
\newcommand{\br}{\mbox{{\bf R}}}
\newcommand{\bz}{\mbox{{\bf Z}}}
\newcommand{\bq}{\mbox{{\bf Q}}}
\newcommand{\bn}{\mbox{{\bf N}}}
\newcommand {\eqn}[1]{(\ref{#1})}

\newcommand{\cp}{\mbox{{\bf P}}^1}
\newcommand{\n}{\mbox{{\bf n}}}
\newcommand{\sbz}{\msc{{\bf Z}}}
\newcommand{\sn}{\msc{{\bf n}}}

\newcommand{\be}{\begin{equation}}\newcommand{\ee}{\end{equation}}
\newcommand{\bea}{\begin{eqnarray}} \newcommand{\eea}{\end{eqnarray}}
\newcommand{\ba}[1]{\begin{array}{#1}} \newcommand{\ea}{\end{array}}

\newcommand{\cleqn}{\setcounter{equation}{0}}
\makeatletter
\@addtoreset{equation}{section}
\def\theequation{\thesection.\arabic{equation}}
\makeatother

\def\npb{Nucl. Phys. {\bf B}}
\def\plb{Phys. Lett. {\bf B}}
\def\mpla{Mod. Phys. {\bf A}}
\def\ijmpa{Intern. J. Mod. Phys. {\bf A}}
\def\cmp{Comm. Math. Phys.}
\def\prd{Phys. Rev. {\bf D}}

\def\vu{\vec u}
\def\vs{\vec s}
\def\vv{\vec v}
\def\vt{\vec t}
\def\vn{\vec n}
\def\ve{\vec e}
\def\vp{\vec p}
\def\vk{\vec k}
\def\vx{\vec x}
\def\vz{\vec z}
\def\vy{\vec y}
\def\vxi{\vec\xi}


\begin{flushright}
La Plata Th-01/07\\October
 2001
\end{flushright}

\bigskip

\begin{center}

{\Large\bf A note on the non-commutative Chern-Simons model on
manifolds with boundary}
\footnote{This work was partially supported by CONICET, Argentina
}

\bigskip
\bigskip

{\it \large Adri\'{a}n R. Lugo} \footnote{ {\sf
lugo@obelix.fisica.unlp.edu.ar} }
\bigskip

{\it Departamento de F\'\i sica, Facultad de Ciencias Exactas \\
Universidad Nacional de La Plata\\ C.C. 67, (1900) La Plata,
Argentina}
\bigskip
\bigskip

\end{center}

\begin{abstract}

We study field theories defined in regions of the spatial non-commutative (NC) plane with a boundary present
delimiting them, concentrating in particular on the $U(1)$ NC Chern-Simons theory on the upper half plane.
We find that classical consistency and gauge invariance lead necessary to the introduction of $K_0$-space of square
integrable functions null together with all their derivatives at the origin.
Furthermore the requirement of closure of $K_0$ under the $*$-product leads to the introduction of a novel notion
of the $*$-product itself in regions where a boundary is present,
that in turn yields the complexification of the gauge group and to consider chiral waves in one sense or other.
The canonical quantization of the theory is sketched identifying the physical states and the physical operators.
These last ones include  ordinary NC Wilson lines starting and ending on the boundary that yield correlation
functions depending on points on the one-dimensional boundary.
We finally extend the definition of the $*$-product to a strip and comment on possible relevance of these results to
finite Quantum Hall systems.

\end{abstract}
\bigskip

\section{Introduction}
\cleqn

In the last years many efforts were put in the study of NC theories since the comparison of NC
geometry on the torus in compactifications of $M$-theory matrix models \cite{cds} and NC gauge
theories as effective actions for string theories with a constant $B$-field \cite{seiwit}.
NC scalar theories were considered in \cite{minraamsei}; NC Wess-Zumino-Witten (WZW)
models in \cite{wzw}.

A novel application to condensed matter systems was recently pursued with the introduction of
non-commutative versions of the Chern-Simons (CS) theory as suitable effective theories of
Fractional Quantum Hall (FQH) fluids.
In reference \cite{suss} Susskind introduced a matrix model version of NC CS theory describing
infinite systems; later in \cite{poly} (see also \cite{raam}) Polychronakos  modified  the
Susskind action by adding a boson field that attaches flux to the electrons and a harmonic
oscillator potential to confine them in such a way to make the theory compatible with the
finiteness of the number of electrons.
Although in any case some properties of the FQH fluids are well reproduced (quantization of the
filling fraction, existence of Laughlin quasi-particles of fractional statistics, etc.) is still
necessary to  check that these descriptions are the right ones (by computing for example
density-density correlations as correlations of operators to be identified).
We believe however that the natural way to study the long-range behaviour of a system with a
finite number of degrees of freedom confined in a finite region of the space is to consider a
(field) theory on that region.
In this direction points out this note, addressing the non trivial problem of
defining NC field theories in spaces different from $\Re^d$ or compactifications thereof, see also recent work in
reference \cite{stern} (other topics related to NC CS theories in infinite space are addressed in \cite{sheikh}).

We start remembering that the straight way to define a NC field theory is to introduce
the usual Groenewold associative $*$-product
\be
f*g\,(\vx) \equiv \exp\left(\frac{i}{2}\, \theta^{\mu\nu} \frac{\partial}{\partial
y^\mu} \frac{\partial}{\partial z^\nu} \right)\; f(\vy)\; g(\vz)\;
\left|_{\vy=\vz=\vx}\right. = \overline{{\bar g}*{\bar f}}(\vx)\label{groene}
\ee
that yields the Moyal bracket and anti-bracket
\bea [f,g]_*(\vx) &\equiv& f*g\, (\vx) - g*f\, (\vx) =
i\,\theta^{\mu\nu}\;\partial_\mu J_\nu[f;g]\cr
J_\mu [f;g] &=& \sum_{m\geq 0}\;
\frac{(-)^m\, 2^{-2m-1}}{(2m+1)!}\; \theta^{\mu_1\nu_1}\dots
\theta^{\mu_{2m}\nu_{2m}}\;
\partial_{\mu_1}\dots\partial_{\mu_{2m}} f(\vx)\; \partial_\mu\,
\partial_{\nu_1}\dots\partial_{\nu_{2m}} g(\vx)\cr
&-& (f\leftrightarrow g)\label{moyalbra}
\eea
\bea
\{f,g\}_*(\vx) &\equiv& f*g\, (\vx) + g*f\, (\vx) = 2\, f(\vx)\,g(\vx) +
\theta^{\mu\nu}\;\theta^{\rho\sigma}\;
\partial_\mu\partial_\rho J_{\nu\sigma}[f;g]\cr
J_{\mu\nu}[f;g] &=& \sum_{m\geq 0}\; \frac{(-)^{m+1}\,2^{-2m-2}}{(2m+2)!}\,
\theta^{\mu_1\nu_1}\dots \theta^{\mu_{2m}\nu_{2m}}\;
\partial_{\mu_1}\dots\partial_{\mu_{2m}} f(\vx)\; \partial_\mu\;\partial_\nu\;
\partial_{\nu_1}\dots\partial_{\nu_{2m}} g(\vx)\cr
&+& (f\leftrightarrow g)\label{moyalantibra}
\eea
These definitions, being local but non-covariant in general, have sense in coordinate
systems pre-determined; working in infinite euclidean space  there is no doubt about which
they are.
Then to get the NC version of the action that defines the theory under consideration we simply
replace the usual field products by $*$-products.
However if the manifold is not $\Re^d$ or if it is a bounded region of it neither the definition
of the $*$-product nor the field theory itself are straightforward.
Typical terms coming from adjoint actions
\bea
e_*^{X}*O*e_*^{-X} &=& O + i\,\theta^{\rho\sigma}\;\partial_\rho J_\sigma[X;\hat A_X(O)]\cr
\hat A_X(O) &\equiv& \sum_{m\geq 0}\; \frac{1}{(m+1)!}\;\underbrace{[X;\dots[X}_m; O]_*\dots]_*
\label{adjbt}
\eea
can not be loosely discarded and as a consequence the (classical) operator formulation
considered i.e. in \cite{gross},\cite{poly} has not an obvious extension either.

For definiteness we will take the space-time of the form $\Sigma= D\times\Re$ with
$D\subset\Re^2$ being the space bounded by $S=\partial D$ and $\Re$ the time axis parameterized
by $x^0\equiv t$.
Furthermore we will restrict the non-commutativity to the space coordinates,
$\theta^{0i}=0;\;\; \theta^{ij}=\theta\;\epsilon^{ij},\;\; i,j=1,2\,$, in some coordinates
$\vx=(x^1 , x^2)$ in $D$ to be specified.
\bigskip

\section{The model}
\cleqn
Let us start by introducing a NC $U(1)$ gauge field $A$ with field strength
\footnote{
In what follows ``$c$" and ``$p$" stand for cyclic permutations and general permutations
including sign respectively while $R(g)\equiv dg*g^{-1}$ and $L(g)\equiv g^{-1}*d g$ will denote
the right and left-invariant Maurer-Cartan forms.
}
\bea
F[A] &\equiv& dA + i\; A\stackrel{\wedge}{*} A =
\frac{1}{2}\; F_{\mu\nu}[A]\; dx^\mu\wedge dx^\nu\cr
F_{\mu\nu}[A] &=& \partial_\mu A_\nu-\partial_\nu A_\mu + i\;[A_\mu;A_\nu]_*\cr
0 &=& \partial_\mu F_{\nu\rho} + i\;[A_\mu ;F_{\nu\rho}]_* + c(\mu\nu\rho)
\eea
where the last line is the NC version of the Bianchi identity; extension of what follows to
$U(N)$ is straightforward introducing Lie algebra generators and suitable traces.
A gauge transformation by the element $g(x)=e_*^{i\phi(x)}= 1 + i\,\phi(x) - \frac{1}{2}\phi*\phi(x) +\dots,$ is given by
\bea
^gA &=& g*A*g^{-1} + i\; R(g)\cr F[^gA] &=& g*F[A]*g^{-1}\label{gauge}
\eea
On a field transforming in the adjoint representation $^g\Phi =
g*\Phi*g^{-1}$ the covariant derivative is defined by
\bea
D_\mu\Phi &\equiv&\partial_\mu\Phi + i\; [A_\mu;\Phi]_*\cr
[D_\mu;D_\nu] \Phi &=& i\;[F_{\mu\nu};\Phi]_*
\eea

Now following the route traced in the introduction let us consider the NC $U(1)$ CS
theory defined by the following action
\bea
S[A] &=& -\frac{k}{4\pi}\;\int_\Sigma\;
L[A]\cr L[A] &=& \frac{1}{2}\left( A\stackrel{\wedge}{*} F[A] + F[A]\stackrel{\wedge}{*}
A \right) - \frac{i}{3}\; A\stackrel{\wedge}{*} A\stackrel{\wedge}{*} A\cr
&=&\frac{1}{4}\; d^3 x\; \epsilon^{\mu\nu\rho}\; \{A_\rho; F_{\mu\nu}
-\frac{i}{3}\;[A_\mu;A_\nu]_*\}_*\label{accion}
\eea
We note that in the commutative case the first term of the lagrangean is simply $A\wedge F[A]$;
in the NC case reality of the action for $A$ real (which becomes manifest in the last line) lead us to
replace it by the symmetric parenthesis.

Several things of the action (\ref{accion}) are worth to be analyzed.
The first one concerns the differential of the lagrangean density; a quick computation shows
\bea
dL[A] &=& F[A]\stackrel{\wedge}{*} F[A]  + \frac{1}{4!}\; dx^\mu \wedge dx^\nu  \wedge
dx^\rho\wedge dx^\sigma\; A^{(4)}_{\mu\nu\rho\sigma}\cr
A^{(4)}_{\mu\nu\rho\sigma} &=& \frac{1}{2}\; [A_\mu ;A_\nu*A_\rho*A_\sigma - i\,\frac{2}{3}\;
\{A_\nu ; \partial_\rho A_\sigma\}_*]_* + p(\mu\nu\rho\sigma)
\eea
In the commutative limit the four form $A^{(4)}$ is identically null (in the non abelian case
too, due to a trace present) leading to the well known, gauge invariant and topological defining expression
for the CS density.
This is not so in the NC case under consideration; however this should not be seen as a
great problem, after all the introduction of the $*$-product breaks general covariance.

What it is certainly relevant for the consistency of the theory has to do with the equations of
motion of the theory.
An arbitrary variation of $S[A]$ under $A \rightarrow A+\delta A$ yields
\be
\delta S[A] = -\frac{k}{4\pi}\; \int_\Sigma\; d^3\vx\;\epsilon^{\mu\nu\rho}\;
F_{\mu\nu}\; \delta A_\rho - \frac{k}{4\pi}\; \int_{\partial\Sigma}\; \delta
B\label{varaction} \ee where the two-form $\delta B = \frac{1}{2}\, \delta
B_{\mu\nu}\,dx^\mu\wedge dx^\nu$ is given by \bea \delta B_{\mu\nu} &=& \delta
A_\mu\;A_\nu -\frac{1}{3}\,\theta^{\rho\sigma}\; \left(J_\sigma[ A_\mu; \{\delta A_\nu
; A_\rho\}_* ] + c(\mu\nu\rho) \right)\cr &+&
\frac{1}{2}\,\theta^{\alpha\beta}\;\theta^{\gamma\delta}\;\partial_\gamma
\left(J_{\beta\delta}[ F_{\mu\nu}; \delta A_\alpha] + 2\, J_{\beta\delta}[
F_{\alpha\mu}; \delta A_\nu] + \partial_\alpha J_{\beta\delta}[\delta A_\mu;
A_\nu]\right) - (\mu\leftrightarrow\nu)\cr & & \eea The third fact concerns  the
crucial gauge invariance of the theory that we like to conserve; we get for the
variation of the action
(\ref{accion})
\bea
S[^g{}A] &=& S[A] + 2\,\pi\, k\; \nu[g] - \frac{k}{4\pi}\; \int_{\partial\Sigma}\; B\cr
\nu[g] &=& \frac{1}{24\,\pi^2}\;\int_\Sigma\;
R(g)\stackrel{\wedge}{*} R(g)\stackrel{\wedge}{*}R(g)\label{gaugevar}
\eea
where the two-form $B = \frac{1}{2}\, B_{\mu\nu}\,dx^\mu\wedge dx^\nu$ is given by
\bea
B_{\mu\nu} &=& \lambda_{\mu\nu}-\epsilon_{\mu\nu\rho}\;\theta^{\rho\sigma}\;
J_\sigma [\phi; \hat A_{i\phi}(L[A]+d\lambda)_{012}]\cr
\lambda_{\mu\nu} &=& A_\mu \; i\,L_\nu
- \frac{\theta^{\rho\sigma}}{6}\;\left( J_\sigma[A_\mu;\{A_\nu; i\,L_\rho\}_* ] +
J_\sigma[i\,L_\mu(g);\{i\,L_\nu(g); A_\rho\}_* ] +c(\mu\nu\rho)\right)\cr
&+&\frac{1}{2}\,\theta^{\alpha\beta}\;\theta^{\gamma\delta}\; \partial_\alpha\partial_\gamma
J_{\beta\delta}[A_\mu;iL_\nu]- \left(\mu\leftrightarrow\nu\right)
\eea
On the other hand $\nu[g]$ is the winding number labeling the homotopy class the map $g$
belongs to.
Then invariance under large gauge transformations imposes the quantization of $k$.
In any case the variation of $\nu[g]$ under a continuous deformation
$g\rightarrow g+\delta g$ is given by
\bea
\delta\nu[g] &=& \frac{1}{24\pi^2}\;\int_{\partial\Sigma}\; H\cr
H &=& \delta g*g^{-1}*dR(g) + dR(g)*\delta g*g^{-1}\; - R(g)\;\stackrel{\wedge}{*}\delta
g*g^{-1}*R(g)\cr
&+& \frac{1}{2}\, dx^\mu\wedge dx^\nu\; \theta^{\rho\sigma}\;
\left( J_\sigma[i\,\delta g\,g^{-1}; R_\mu*R_\nu*R_\rho]  +
J_\sigma[R_\rho*i\,\delta g\,g^{-1}; R_\mu*R_\nu]\right.\cr
&+&\left. p(\mu\nu\rho)\right)\label{deltanu}
\eea
If $\Sigma$ has no boundary the variation is automatically zero; this also holds if $g$ obeys
suitable boundary conditions.
In the commutative context the boundary term drops out by imposing Dirichlet boundary conditions
on just {\it one} gauge field component on the boundary (i.e $A_\sigma$); consistently the good
gauge transformations are those that not depend on the coordinate in that direction on the
boundary ( $R_\sigma|_{\partial\Sigma}=0$).
In this case we get that $\nu[g]$ is invariant under continuous deformations; in
particular it must be zero for $g$ connected with the identity (small gauge transformations).

But in the NC case things changes dramatically; direct inspection to (\ref{varaction}),
(\ref{gaugevar}) and (\ref{deltanu}) reveals that it is absolutely needed to impose
Dirichlet like boundary conditions on {\it two} fields instead that one in such a way that
boundary terms drop out in order to get a (at least classically) consistent, gauge invariant
theory.
But while in the commutative case standard Dirichlet conditions suffice in the NC
case at hand it is not so.
And the reason relies in the $*-$product that involves spatial derivatives of {\it all}
orders at the boundary; this means that is not sufficient to impose that a field be
zero at the boundary, but {\it all} their derivatives too.
Consistently the gauge transformations generated by $g=e_*^{i\,\phi}$ must go to the identity
identically on it in the sense that $\phi$ and all its derivatives must go to zero.
Thus we are enforced to take as boundary conditions the following ones
\be
\partial_1^p\partial_2^q A_1(\vx,t)|_S = \partial_1^p\partial_2^q A_2(\vx,t)|_S =0 \;\;\;,
\;\;\; p,q=0,1,2,\dots\label{bc}
\ee
Certainly these conditions indicates us that near
the boundary the fields can not be analytical. Fortunately spaces of functions of this
kind are known in the mathematical literature \cite{mage}.
By the meantime we just accept we are working on this kind of spaces that generically we will
denote as $C(D)$; we will work out some examples in Section $4$ while in Section $5$ we
will address the highly non trivial problem concerning its closure under the $*$-product.

\section{Canonical quantization, physical states and operators}
\cleqn
Once we have the setting outlined in the past section we can rewrite the action (\ref{accion})
as
\be
S[A] = \frac{k}{4\pi}\; \int\;dt \int_D d^2\vec x\; \epsilon^{ij}\;\left(
A_i\; \partial_0 A_j -\;A_0\;F_{ij}\right)\label{accionef}
\ee
Now we can proceed to fix the gauge; as usual in CS context we will fix the temporal gauge
$A_0=0$.
Then the equation of motion for the spatial components of $A$ are simply
\be
\partial_0 A_i(t,\vec x) = 0 \;\;\; \longrightarrow\;\;\; A_i = A_i(\vx)
\ee
Since now on we will think about $A$ as a connection one-form on $D$ obeying
(\ref{bc}).
On the other hand the equation of motion for $A_0$ gives the constraint
\be
F_{12}= \partial_1 A_2-\partial_2A_1 + i[A_1,A_2]_* =0 \label{const}
\ee
classically solved by $A_i (\vx) = i\,\partial_ig*g^{-1}(\vx)$.
Time-independent gauge transformations (with the boundary conditions discussed) are left
over as usual.
However we would like to stress a radical difference with the
commutative case through a concrete example.
By taking a disk as space and having standard Dirichlet boundary conditions time independent
gauge transformations are to be interpreted as global symmetries, the loop group in this case,
and the states of the theory lie in representations of them instead of being invariant
under such transformations \cite{cscq}.
This fact becomes evident by integrating out the field $A_0$ (cf.(\ref{accionef}));
we finish with a chiral WZW model in the group variable solution of the constraint
(\ref{const}) that naturally realize such a symmetry, a chiral Kac-Moody algebra.
In the NC case formally we can make the same yielding a NC version of the chiral WZW alluded
to \cite{bepis}.
However the strong boundary conditions imposed make trivial this action
(the loop group has disappeared!) and we must conclude that quantum mechanically we have
to impose (\ref{const}) as a constraint on the Hilbert space of the theory,
\be
\hat Q(\phi)\, |phys> = 0 \label{phystat}
\ee
In this sense the quantization resemble the boundary-less case \cite{cscq}.

Going to the quantization of the theory we follow \cite{wit} and read the quantum commutators
between field operators from the symplectic structure inherited from (\ref{accionef})
\be
[\hat A_i(\vx),\hat A_j(\vy)] = i\,\frac{2\pi}{k}\;\epsilon_{ij}\;\delta^2(\vx-\vy)
\label{commut}
\ee
Thus the generator of infinitesimal residual gauge transformations with element
$g=e_*^{i\,\phi}$ is given by
\be
\hat Q (\phi) = \frac{k}{2\pi}\;\int_D\; d^2\vx\; \phi(\vx)\; \hat F_{12} \ee where the
operator $\hat F_{12}$ is as in (\ref{const}) with the replacement with the $\hat
A$-operator and a suitable normal ordering.
In fact it verifies the following algebra
\be
[\hat Q(\phi_1),\hat Q(\phi_2)] = -\hat Q([\phi_1,\phi_2]_*)\label{qalg}
\ee
that show  that the operator representing in  Hilbert space the NC $U(1)$ group is
\be
\hat U(g)\equiv \exp(i\hat Q(\phi))  \;\;\;,\;\;\;
\hat U(g_1)\;\hat U(g_2)\equiv \hat U(g_1*g_2)
\ee
Furthermore it generates the infinitesimal residual gauge transformations
\bea
[i\hat Q(\phi), \hat A_i(\vx)] &=& -\left(\partial_i\phi + i\;
[\hat A_i,\phi]_*(\vx)\right)\cr
[i\hat Q(\phi), \hat F_{12}(\vx)] &=& [i\,\phi , \hat F_{12} ]_*(\vx)
\eea
from where it follows
\bea
\hat U(g)\; \hat A_i(\vx)\; \hat U(g)^{-1} &=&  g*\hat A_i(\vx)*g^{-1} + i\;R(g)\cr
\hat U(g)\; \hat F_{12}(\vx)\;\hat U(g)^{-1} &=&  g*\hat F_{12}(\vx)*g^{-1}
\eea
Imposing gauge invariance means searching for states invariant under $\hat U(g)$;
under small gauge transformation this is no other condition that the Gauss law
(\ref{const}); large gauge transformations instead corresponds to non trivial $\phi$.

On the other hand physical operators must be the gauge invariant ones. In commutative gauge theories the natural
local operators are those constructed from powers of the field strength transforming in the adjoint
representation; the non-local ones are the Wilson loops. In NC gauge theories things are more involved; local
operators do not seem to exist, the problem is addressed in references \cite{ishi}, \cite{gross}. Let us review
what we need here to guess the physical operators in our context. First let us introduce a charge-$q$ open Wilson
line from the base point $\vx$, the reference point to compute the $*$-product, to $\vx + \Delta\vx$ on the path
$\gamma$ parameterized by $\vx(s) = \vx + \vxi(s)\;,\;\vxi(0)=\vec 0 \;,\; \vxi(1) = \Delta\vx$ \bea \hat
W_q[\gamma] &\equiv& P_*\; \exp{\left(i\,q \int_0^1 \; ds\; \dot{\xi}^i(s)\;\hat A_i(\vx + \vxi(s))\right)}\cr &=&
\sum_{n=0}^{\infty} (i\,q)^n\; \int_0^1\; ds_1 \int_{s_1}^1\; ds_2\dots \int_{s_{n-1}}^1\;
ds_n\;\dot{\xi}^{i_1}(s_1)\;\dots \dot{\xi}^{i_n}(s_n)\cr &\times& \hat A_{i_1}(\vx + \vxi(s_1))*\dots *\hat
A_{i_n}(\vx + \vxi(s_n))\label{wilson} \eea Under a gauge transformation the $q=1$ operator goes to
\be
\hat U(g)\;\hat W_1[\gamma]\;\hat U(g)^{-1}= e_*^{i\,\phi(\vx)}*\hat W_1[\gamma]*
e_*^{-i\,\phi(\vx +\Delta\vx)}
\ee
In commutative gauge theories we get gauge invariance just closing the line to make a loop.
But in NC gauge theories it is necessary to introduce a kind of Fourier
transform; specifically if $\hat O(\vx)$ is any local operator transforming in the adjoint
representation and the path $\gamma$ is specified by
$\xi^i(s)= -\theta\,\epsilon^{ij}\;k_j\;s\;$ then the set of operators
\be
\hat{\tilde O}(\vk;\gamma) = \int_D \;d^2\vx\;\hat O(\vx)*\hat W_1[\gamma] *e^{i\vk\cdot\vx}\label{bulk}
\ee
is gauge invariant \cite{ishi}, \cite{gross}.
This property heavily relies on the capability of getting out boundary terms (c.f.
(\ref{adjbt})) ; in infinite space this is natural from the usual boundary conditions at
infinity, here from the boundary conditions imposed on $S=\partial D$.
But due to the no existence in general of translation invariance and the non local character
of the $*$-product we presume that not all of the operators defined in (\ref{bulk}) will be well
defined; we will call them ``bulk operators".
However we have at hand in our context another interesting set of ``boundary operators";
in fact if we take the endpoints of the path $\gamma$ to lie on the boundary, then
automatically the Wilson lines (\ref{wilson}) are gauge invariant.
\footnote{
We might think that arbitrary small loops starting and ending at the same point
on the boundary would have a local character; however this is not so because it is not
possible to make them arbitrary small since they become trivial due to the strong boundary
conditions imposed on the fields.
}

Resuming, we have found that the physical, gauge invariant, correlation functions of
the theory are given by VEV's in the physical states (\ref{phystat}) of strings of
the boundary and bulk operators previously defined.

\section{Explicit construction of basis in $C(D)$}
\cleqn
As seen in Section $2$ we were led to introduce spaces of functions defined on a region of
the plane $D$ which vanish, as well as their derivatives of all order, at the boundary
$\partial D= S$.
We would like to work out in this section explicit examples of basis in such spaces.

Let us start considering the simplest one of such regions, the upper half-plane
$H= \{\vx=(x,y) \in\Re^2 : y>0\}$, with $\partial H $ being the $x$-axis ($y=0$).
Later, through clever (or naive, as you like) mappings, we will be able to define basis in
other regions of potential interest.

Usually, functions forming a basis on $L^2(H)$, the space of  square integrable functions on
$H$, are obtained as solutions of the Laplace equation
$-\partial_i\partial_i\Psi(\vx) = \lambda^2\; \Psi(\vx)$, in virtue of the completeness of the
laplacian operator when Dirichlet or Neumann boundary conditions are considered.
They are written in the form
\be
\Psi_{k,m}(\vx) = \frac{e^{i\,k\, x}}{\sqrt{2\pi}}\; \psi_m(y)\label{base}
\ee
where $k\in\Re$ and $m$ some other label.
Their derivatives of order $(p,q)$ in $x$ and $y$ respectively are obviously given by
\be
\partial_x^p\partial_y^q\Psi_{k,m}(\vx) = (i\,k)^p\;  \frac{e^{i\,k\, x}}{\sqrt{2\pi}}\;
\psi^{q)}_m(y)\label{deri}
\ee
The functions $\psi_m(y)$ will depend on the boundary conditions on the $x$-axis.
If Dirichlet conditions $\Psi_m(x,y=0)=0$ are imposed we can take $\psi_m(y)\sim
\sin m\,y$, if Neumann conditions $\Psi^{1)}_m(x,y=0)=0$ are instead considered then
$\psi_m(y)\sim\cos m\,y$; in both cases $m$ is a {\it continuous}, non-negative number
and $\lambda^2 = k^2 + m^2$.

But what happen if we would like to impose that all the derivatives (\ref{deri}) be null
at the boundary?
We must take $\psi_m(y)$ such that
\be
\psi^{q)}_m(y)|_{y=0} = 0 \;\;\; ,\;\;\; q = 0,1,2,\dots\label{strongbc}
\ee
In other words, we are led to consider a subspace of $L^2(\Re^+)$ that includes only those
functions that are zero together with all their derivatives at the origin.
We denote this space as $K_0$; orthonormal and {\it discrete} basis exist on it
and furthermore it has the at all obvious property of being dense in $L^2(\Re^+)$, that is,
any function in $L^2(\Re^+)$ can be approximated as well as we like by some sequence in
$K_0$ \cite{mage}.
The question is, what are the $\psi_m{}'s$?
Certainly they can not come as solutions of the flat Laplace equation; we do not know either if
there exists some metric on $\Re^+$ from which such a basis could be a complete set of the
associated Laplace operator.
However we will construct them in the following way.
First we introduce the harmonic oscillator wave-functions
\be
\phi_{n}(\xi) = (2^n\, n!\,\sqrt\pi)^{-\frac{1}{2}}\; e^{-\frac{\xi^2}{2}}\;
H_n(\xi)\;\;\; , \;\;\; n=0,1,2,\dots
\ee
As it is well-known they are basis in $L^2(\Re)$ orthonormal wrt the standard scalar product
\be
(\phi_m ,\phi_n)\equiv \int_S\; d\xi\;\overline\phi_{m}(\xi)\;\phi_n(\xi) =
\delta_{m,n}\label{scalarprod}
\ee
where $S=\Re$.
If we would like to restrict $L^2(\Re)$ to $L^2(\Re^+)$ with Dirichlet (Neumann) boundary
condition $\phi(0)=0$ ($\phi'(0)=0$) it would suffice in view of the parity property of the
Hermite polynomials $H_m(-\xi) = (-)^m\, H_m(\xi)$ to take $m$ odd (even).
Then it follows that the set
\be
\{\sqrt{2}\;\phi_{2m +1}\,,\; m=0,1,\dots\}\;\;\; (\{\sqrt{2}\;\phi_{2m}\,,\; m=0,1,\dots\})
\label{base1}
\ee
constitute basis in $L^2(\Re^+)$ orthonormal wrt the scalar product
(\ref{scalarprod}) with $S=\Re^+$.

Now we define the following functions on $\Re^+$
\be
\psi_m (y)\equiv \sqrt{2}\;\frac{\phi_{2m +1}(\frac{1}{y})}{y}=
2^{-m}\, ((2m+1)!\,\sqrt\pi)^{-\frac{1}{2}}\; e^{-\frac{1}{2y^2}}\;
\frac{H_{2m+1}(\frac{1}{y})}{y}\;\;\;,\;\;\;m=0,1,2,\dots \label{base2}
\ee
These functions die as $y^{-2}$ at infinity and as $y^{-2m-2}\,e^{-\frac{1}{2y^2}}$
near $y=0$ obeying therefore (\ref{strongbc}); so they belong to $K_0$.
And what is stronger, they define a mapping in $L^2(\Re^+)$ from the set (\ref{base1})
to (\ref{base2}) that conserve the scalar product
\be
(\psi_m ,\psi_n) = \int_{\Re^+}\; dy\;\overline\psi_{m}(y)\;\psi_n(y) =
\delta_{m,n} =(\sqrt{2}\,\phi_{2m+1} ,\sqrt{2}\,\phi_{2n+1})
\label{scalarprodbis}
\ee
From standard theorems we conclude that (\ref{base2}) is a discrete, orthonormal basis in
$K_0$,  as stated.
Thus we have obtained a basis in $C(H)$ in the form (\ref{base}) obeying
\be
\int_H\; dx\,dy\; \overline\Psi_{k,n}(\vx)\;\Psi_{k',n'}(\vx) = \delta(k-k')\;\delta_{n,n'}
\label{2dscalarprod}
\ee

We can obtain basis in other interesting bounded regions by mappings of $H$ to them. A
first example is a strip extended in the $x$-axis of width $L$ in the $y$-axis.
The mapping in question is very popular for string and conformal field theorists;
if $z=z_1 + i\,z_2 \;\; , \, z_2>0,\;$ is the complex coordinate on the upper half-plane
then it is given by
\be
z = e^{\frac{\pi}{L}(x+i\,y)} \;\;\longleftrightarrow \;\; \left\{
\begin{array}{ll} z_1 = e^{\frac{\pi}{L}\,x}\;\cos\frac{\pi}{L}\,y\\
z_2= e^{\frac{\pi}{L}\,x}\;\sin\frac{\pi}{L}\,y\end{array} \right.\label{cv1}
\ee
where the coordinates $\;-\infty<x=\frac{L}{\pi}\ln|z|<\infty\;\;,\;
0<y=\frac{L}{\pi}\tan^{-1}\frac{z_2}{z_1}<L\;,$ label the strip $S_L$.
Then the functions on the strip defined by
\be
\Phi_{k,n}(\vx)\equiv \frac{\pi}{L}\;e^{\frac{\pi}{L}\,x}\;
\Psi_{k,n}(\vz)|_{\vz(\vx)=(\ref{cv1})}\label{stripfunc}
\ee
are orthonormal in the sense of (\ref{2dscalarprod}) (with $H$ replaced with $S_L$).
And what is important for us, it is straightforward to verify that on the boundary of
$S_L$, i.e. the lines $y=0$ and $y=L$, they verify that their derivatives in $x$, $y$
of any order are null; we conclude that (\ref{stripfunc}) are basis in $C(S_L)$.

Lastly we consider a disk  $D_R$ of radius $R$; we can obtain it from $H$ in the following way. First we
compactify  the $x$-coordinate on a circle of radius $R$ getting a half-cylinder, then we stretch the (positive)
$y$-coordinate from the border to the center of the disk. Specifically if we parameterize the disk by radial and
angular coordinates $(\rho, \varphi)$, then the mapping is defined by
\bea
\varphi &\equiv& \frac{x}{R} \sim
\varphi + 2\pi\cr \rho &\equiv& R\; f(y)\;\;\;,\;\;\; f(0)=1\;\;\;, f(\infty) = 0\label{mapdisk} \eea
We take $f'(y)<0$ for any $y>0$, and going to zero at infinite as a power lesser that $\frac{3}{2}$ in order for
to be regular at the origin.
As before with the strip we can introduce the functions and orthogonality relations
\bea \Theta_{mn}(\rho,\varphi) &\equiv& \frac{e^{i\,m\,\varphi}}{\sqrt{2\pi}}\;
\frac{\psi_n(y)}{\sqrt{-\rho(y)\,\rho'(y)}}\,|_{y(\rho)=(\ref{mapdisk})}\cr \delta_{m,m'}\;\delta_{n,n'}&=&
\int_0^R\;\int_0^{2\pi}\;d\rho\;\rho\;d\varphi\; \overline\Theta_{mn}(\rho,\varphi)\;
\Theta_{m'n'}(\rho,\varphi)\label{teta} \eea where the momenta are now quantized $k=\frac{m}{R},\, m\in Z$. They
verify that their derivatives of any order on $\partial D_R \sim S^1$ are null, then they expand $C(D_R)$.


\section{About the closure of $C(D)$ under the $*$-product}
\cleqn

From the well-known smearing of localized functions under the $*$-product in $\Re^d$ which makes manifest its
non-local character (see for example \cite{nekra}) a very important question that rises in our framework
is if the $*$-product of two functions in $C(D)$  belongs to $C(D)$.
It is clear that this property of closure is crucial for the consistency of the whole construction carried out so
far, so here we address this problem; for sake of definiteness we will consider $C(H)$ and the basis
defined in (\ref{base}), (\ref{base2}).
\footnote{
Indeed all we need is a basis obeying (\ref{strongbc}), not necessarily that constructed in (\ref{base2}).
}

The $*$-product of two functions of the basis can be easily worked out with the result
\be
\Psi_{k,m}*\Psi_{k',m'}(\vx) = \e^{\frac{\theta}{2}\,k'\,\partial_y}\,\Psi_{k,m}(\vx)\;
 \e^{-\frac{\theta}{2}\,k\,\partial_y}\,\Psi_{k',m'}(\vx)\label{product}
\ee The question to be answered is then, what is the value of the rhs of (\ref{product}) (and that of its
derivatives of any order), in particular at $y=0$? The translation formula
\be
\e^{a\,\partial_y}f(y)=f(y+a) \;\;\;\;,\;\;\; a \in \mathcal{C}\label{trans}
\ee
is naturally valid for any $f$ {\it analytical} in $y$.
But our functions (\ref{base2}) are {\it not} analytical in $y=0$ so another route must be
followed to {\it define} the lhs of (\ref{trans}) and consequently the $*$-product.

In order to guess the right definition we observe two facts,
\begin{itemize}
\item We need to have defined $\psi_m(y)$ only for $y>0$;
\item A translation passing through the essential singularity at $y=0$ is not possible.
\end{itemize}
Accordingly, we first define an extension of the basis (\ref{base2}) to the whole line as follows
\be
\tilde\psi_m(y)\equiv
\lim_{\epsilon\rightarrow0^+}\; \frac{1}{2\pi i}\; \oint_{C_\epsilon}\; dw\;\frac{\psi_m(-i w)}{w - i y}
= \left\{ \begin{array}{ll} \psi_m(y)&,\;\;\; y>0 \\0&,\;\;\; y<0\end{array}\right.
\label{psidef}
\ee
where $C_\epsilon$ is the contour that encloses the semi-annulus centered at the origin
of minor radius $\epsilon>0$ and major radius $\infty$  in the upper half-plane.
Therefore in the whole plane (but effectively no null in $H$) we will consider the set
\be
\tilde\Psi_{km}(\vx) \equiv \frac{\e^{ikx}}{\sqrt{2\pi}}\; \tilde\psi_m(y)
\ee
Equation (\ref{psidef}) is just the continuous gluing at $y=0$, together with the derivatives
of {\it any} order, of the function $\psi_m$ for $y>0$ with the function null for $y<0$.
In other words, $\tilde\psi_m\in C^\infty(\Re)$.

We are then led to introduce  the following  action of the translation operator
\be
\e^{a\,\partial_y}\;\tilde\psi_m(y)\; =
\left\{ \begin{array}{ll} \psi_m(y+a)&,\;\;\; y>M(0,-a) \\0&,\;\;\; y<M(0,-a) \end{array}\right.
\ee
where we denote $\,M(a,b,\dots)= \it{maximum}(a,b,\dots), m(a,b,\dots)= \it{minimum}(a,b,\dots)$.
The $*$-product implied by this definition is
\be
\tilde\Psi_{k,m}*\tilde\Psi_{k',m'}(\vx) =\left\{ \begin{array}{ll}
\Psi_{k,m}(x,y + \frac{\theta}{2}\,k')\;\Psi_{k',m'}(x, y-\frac{\theta}{2}\,k)
&,\;\;\; y>\frac{\theta}{2}\, M(0,k,-k') \\0 &,\;\;\; y<\frac{\theta}{2}\, M(0,k,-k')
\end{array}\right.\label{starprod}
\ee
It can be checked by direct inspection that associativity holds for it.
Moreover, straightforward analysis of the rhs of (\ref{starprod}) yields the conclusion
that it defines a  $C^\infty$-function in both variables  {\it except} in the case
$k<0, k'>0$ where a discontinuity at $y=0$ is present.
In all the other cases it can be succinctly written as
\be
\tilde\Psi_{k,m}*\tilde\Psi_{k',m'}(\vx) =
\tilde\Psi_{k,m}(x,y + \frac{\theta}{2}\,k')\;\tilde\Psi_{k',m'}(x,y-\frac{\theta}{2}\,k)
\label{starprodpart}
\ee
from where it is evident that if we take all the momenta to be non-negative {\it or}
all to be non-positive, then the rhs of (\ref{starprodpart}) evaluated at $y=0$ is zero;
the same happens with the $y$-derivatives of all orders.
In other words, we arrived to the result that the $*$-product (\ref{starprodpart}) closes in $C(H)$ if
we restrict the space of functions to waves propagating in the $x$-direction
to the right {\it or} to the left; we can not mix both of them consistently.
\footnote{
Note that the momentum of the wave in the $x$-variable of the rhs of (\ref{starprod}) is
$k+k'$ that is positive (negative) if both $k,k'$ are positive (negative).
}
From sake of completeness we quote another standard way of writing (\ref{starprodpart})
through the distribution $K(\vx_1,\vx_2;\vx)$; for $k\geq0$ we have
\bea
f*g(\vx) &=& \int_H\; d^2\vx_1\;d^2\vx_2\; K(\vx_1,\vx_2;\vx)\; f(\vx_1)\;g(\vx_2)\cr
K(\vx_1,\vx_2;\vx) &=& \frac{1}{(\pi\theta)^2}\; \e^{i\,\frac{2}{\theta}(\vx-\vx_1)\times(\vx-\vx_2)}\;
\left\{\begin{array}{ll}1 &,\;\;\;  0< y_2<y<y_1  \\0 &,\;\;\; otherwise\end{array}\right.
\label{starprodpartbis}
\eea
The property of closure is manifest from this definition; on the other hand
associativity is equivalent to the identity
\be
\int_H\; d^2\vx_4\;K(\vx_1,\vx_2;\vx_4)\;K(\vx_4,\vx_3;\vx) = \int_H\;
d^2\vx_4\;K(\vx_2,\vx_3;\vx_4)\;K(\vx_1,\vx_4;\vx) \ee It can be easily verified that
both sides give the expression
\be
\frac{ \e^{ i\frac{2}{\theta}\left((x-x_1)y_1 +(x_1-x_3)y_2+(x_3-x)y_3\right) } }
{i\, 2\, \pi^3\,\theta^2\,\Delta x}\;
\delta(\Delta y)\;\Theta(m(y_1-y_2,y_2-y_3))\;\left(
\e^{i\frac{2}{\theta}\,\Delta x\, m(y_1-y_2,y_2-y_3)}-1\right)
\ee
where $\Delta x\equiv x_1-x_2+x_3-x  ,\,\Delta y\equiv y_1-y_2+y_3-y\,$, and $\Theta(x)$ is
the step function.

The rather surprising result (at least for the author) that we have found concerning the
chirality of the fields has a very important consequence in the case of the NC CS theory
studied in this paper.
Because of the impossibility of expanding real fields in terms of chiral waves,
we are led to {\it complexify} the theory, i.e. to think about $U(1)^c$ rather than
$U(1)$ as the gauge group and consequently all the components of the gauge fields
(as well as the field $\phi$ that define an element of the gauge group) as complex fields
of definite chirality in the sense introduced above.

\section{Conclusions and Outlook}
\cleqn

We have consider in this letter the problem of formulating NC CS field theories on
finite regions of space.
We found in first instance that consistency and gauge invariance impose very strict and unusual
boundary conditions on the fields.
\footnote{
It is worth to note to this respect that a free scalar field theory under these boundary conditions
trivializes in the sense that there do not exist classical solutions other than the null one.
}
Furthermore the closure of the $*$-product in the space of functions considered led to
the needed of complexify the gauge group and restrict the space of functions to waves
of definite chirality.
We show  that in the CS case it is possible to go ahead and quantize consistently the theory in
a canonical way.
The boundary effects as well as the NC character of the theory are encoded in the
space of functions to be considered {\it and} in the observable ``bulk" and ``boundary"
operators defined in Section $3$; the correlation functions of the last ones depending on
points on the one-dimensional boundary.
We have not addressed here the problem of finding the physical states (presumably only one if $D$ is topologically
trivial); a second step should be the computation of correlation functions.

A possible application it should be worth to investigate is in the context of the Fractional Quantum Hall effect.
In fact another hint in this direction (other than previous work on the subject) relies in the following observation:
it is not difficult to extend the definition (\ref{starprodpartbis}) to a {\it strip} defined
by $S=\{(x,y) : x\in\Re,  a<y<b\}$ in the following way
\be
f*g (\vx) \equiv \frac{1}{(\pi\,\theta)^2}\;\int_0^{b-y}\,dy_1\;\int_0^{y-a}\,dy_2\; e^{i\frac{2}{\theta}\,x\,(y_1+y_2)}
\; \tilde f (\frac{2\,y_2}{\theta};y+y_1)\; \tilde g (\frac{2\,y_1}{\theta};y-y_2)\label{starprodstrip}
\ee
where $\tilde f(k;y)$ is the Fourier transform of $f(\vx)$ in the $x$-variable
\be
\tilde f(k;y) \equiv \int_\Re\; dx\; e^{-i\,k\,x}\; f(\vx)
\ee
The $*$-product (\ref{starprodstrip}) is associative and non-commutative; it can be equally defined by a nucleus
$K(\vx_1,\vx_2;\vx)$ like that in (\ref{starprodpartbis}) with the restriction $\,a<y_2<y<y_1<b$.
Furthermore it is easily verified that it closes in the space of functions null together with their derivatives of any
order in $y=a$ and $y=b$ (the boundary of the strip).
What is more, it has the surprising property of closing not just on functions with positive momenta like
(\ref{starprodpartbis}) made, but in the space of functions with momenta restricted to the region
$0\leq k\leq \frac{2(b-a)}{\theta}$.
This is exactly the region of allowed momenta inside each Landau level in the treatment of the Quantum Hall effect on
the strip \cite{girvin} that leads to a superficial density of states equal to $\frac{2}{\theta}$
(or equivalently, to the quantization of the area of the strip in units of $\frac{\theta}{2}$);
here it emerges in a natural way from the structure of the $*$-product (\ref{starprodstrip}).
We ask if, among other things,  it would be possible to interpret in the Fractional Quantum Hall context the bulk and
boundary operators as related to bulk and chiral edges states, and the correlation functions with Laughlin wave functions.
All these issues are under current investigation \cite{next}.
\section*{Acknowledgements}

We would like to thank Daniel Cabra, Guillermo Silva, Nicol\'{a}s Grandi, Enrique Moreno, and Fidel Schaposnik
for discussions, Mario Rocca for references pointed out and specially  Horacio Falomir for invaluable support
in Section $4$.




\begin{thebibliography}{99}

\bibitem{cds} A. Connes, M. Douglas and A. Schwarz, ``Non commutative geometry and
matrix theory: compactification on tori", hep-th/9711162.

\bibitem{seiwit} N. Seiberg and E. Witten, JHEP {\bf 09} (1999), 32.

\bibitem{minraamsei} S. Minwalla, M. Van Raamsdonk and N. Seiberg,
``Non commutative perturbative dynamics", hep-th/9912072;
M. Van Raamsdonk and N. Seiberg: ``Comments on non commutative perturbative dynamics",
hep-th/0002186.

\bibitem{wzw}
E. Moreno and F. Schaposnik, ``The Wess-Zumino-Witten term in noncommutative two
dimensional fermion models", hep-th/0002236;
K. Furuta and T. Inami, ``Ultraviolet property of noncommutative Wess-Zumino-Witten model",
hep-th/0004024;
E. Moreno and F. Schaposnik, ``Wess-Zumino-Witten and fermion models in noncommutative space",
hep-th/0008118;
A. Ghezelbash and S. Parvizi, \npb 592 (2001), 408, hep-th/0008120;
A. Lugo, ``Correlation functions in the non-commutative WZW model", \plb 511 (2001), 101,
hep-th/0012268.

\bibitem{suss} L. Susskind, ``The quantum Hall fluid and the non-commutative Chern-Simons
theory", hep-th/0101029.

\bibitem{ishi} N. Ishibashi, S. Iso, H. Kawai and Y. Kitazawa, ``Wilson loops in
non-commutative Yang-Mills", hep-th/9910004.

\bibitem{poly} A. Polychronakos, ``Quantum Hall states as matrix Chern-Simons theory",
hep-th/0103013.

\bibitem{stern} A. Pinzul and A. Stern, ``Absence of the Holographic Principle in Noncommutative Chern-Simons
Theory",  hep-th/0107179.

\bibitem{sheikh} M. Sheikh-Jabbari, hep-th/0102092; A. Das and M. Sheikh-Jabbari, hep-th/0103139.

\bibitem{raam} S. Hellerman and M. Van Raamsdonk, ``Quantum Hall physics equals
non commutative field theory", hep-th/0103179.

\bibitem{gross}
S-J. Rey and R. von Unge, hep-th/0007089;
S. Das and S-J. Rey, hep-th/0008042;
D. Gross, A. Hashimoto and N. Itzhaki, hep-th/0008075.

\bibitem{cscq}
E. Witten, \cmp\, 121 (1989), 351;
S. Elitzur, G. Moore, A. Schwimmer and N. Seiberg, \npb 326 (1989), 108;
J. Labastida and A. Ramallo, \plb 227 (1989), 92;
M. Bos and V. Nair, \ijmpa 5 (1990), 959;
A. Lugo, \plb 273 (1991), 231;
R. P. Balachandran, G. Bimonte, K. S. Gupta and A. Stern, Int. J. Mod. Phys. {\bf A7} (1992), 4655 and 5855.

\bibitem{bepis} N. Grandi and G. Silva, \plb 507 (2001), 345, hep-th/0010113.

\bibitem{wit} E. Witten, Comm. Math. Phys. {\bf 92} (1984), 455.

\bibitem{mage} J. Lions and E. Magenes, ``Probl\`emes aux limites non homog\`enes et
applications", vol. 1 and 2, Dunod (Paris), 1968.

\bibitem{nekra} M. Douglas and N. Nekrasov, ``Noncommutative field theory", hep-th/0106048.

\bibitem{girvin} S. Girvin, ``The Quantum Hall Effect: Novel Excitations and Broken Symmetries", cond-mat/9907002.

\bibitem{next} D. Cabra, A. Lugo and G. Silva, work in progress.
















\end{thebibliography}
\end{document}